\documentclass[12pt]{article}

\usepackage{amsfonts}
\usepackage{amsmath}
\usepackage{amscd,amssymb}
\usepackage{hyperref}
\usepackage{graphicx}
\usepackage{amsthm}

\newcommand{\be}{\begin{equation}}
\newcommand{\ee}{\end{equation}}
\newcommand{\bea}{\begin{eqnarray}}
\newcommand{\eea}{\end{eqnarray}}
\let\bm=\bibitem

\def\m{\mu}
\def\n{\nu}

\def\a{\alpha}

\def\del{\partial}
\def\4{{\sst{(4)}}}

\def\f{\phi}
\def\del{\partial}
\def\b{\beta}

\begin{document}
\pagenumbering{roman}

\begin{titlepage}

\vspace{3.0cm}

\

\

\

\centerline{\Large \bf Intersections of S-branes with Waves and Monopoles}
\vspace{1.5cm}

\centerline{Mert Besken$^1$ and Nihat Sadik Deger$^{2}$}

\

\noindent
$^1$ Dept. of Physics and Astronomy, University of California, Los Angeles, CA 90095, USA \\
$^2$ Dept. of Mathematics, Bogazici University, Bebek, 34342, Istanbul-Turkey 

\

\centerline{mbesken@physics.ucla.edu, sadik.deger@boun.edu.tr}

\vspace{1.5cm}

\centerline{\bf ABSTRACT}
\vspace{0.5cm}

We construct intersections of S-branes with waves and Kaluza-Klein monopoles. There are several possible ways to add a monopole
to an S-brane solution similar to p-branes. On the other hand, one may add a wave only to the transverse space of an S-brane 
unlike a p-brane where wave resides on its worldvolume. The metric function of the wave is a
harmonic function of the remaining transverse directions and an extra condition on integration constants is needed. We also show that it is
not possible to add an S-brane to p-brane intersections whose near horizon geometry has an AdS part.

\vspace{2cm}

\end{titlepage}

\pagenumbering{arabic}



\section{Introduction}

S-branes are time dependent solutions of supergravity theories with 
spacelike  worldvolume. They carry electric or magnetic charge like static p-branes but unlike
them they do not preserve any supersymmetry and their transverse space can be spherical or hyperbolic in addition
to flat. As supergravity solutions, they were first obtained in  \cite{pope1, pope2} but the name {\it S-brane} was given in \cite{s1} where 
they were shown to be a type of D-branes that open strings can end with Dirichlet boundary condition applied along 
the time direction. Main motivations to study these objects include dS/CFT correspondence \cite{ds}, tachyon condensation \cite{sen}
and getting accelerating cosmologies from String/M-theory \cite{townsend}-\cite{present}. 

After the study of single S-brane solutions in detail \cite{s2, s3} a natural question to ask was whether they intersect 
like p-branes (see \cite{gaunt} for a review). In  \cite{deger} all intersections among S-branes in D=11 supergravity were obtained and it was shown that 
for every supersymmetric p-brane intersection there is a corresponding S-brane intersection and their metrics obey the {\sl harmonic function rule}
\cite{int1, int2, int3}, 
i.e. worldvolume and transverse directions are multiplied by appropriate powers of metric functions of each brane. Unsurprisingly, this holds in
10-dimensions too \cite{ivas, ohta1}. In these, brane charges are independent and hence a brane can be removed from the solution by setting its charge to zero.
In \cite{nonstandard} new S-brane intersections were obtained by relaxing the independence condition. In \cite{mas}
it was found that S-branes can also intersect with p-branes where the metric functions are multiplications of p- and S-brane metric functions and in all physical fields
radial and time variables are separated.  In the work of \cite{mas} S-brane was located so that its worldvolume included the radial
coordinate of the p-brane. Later in \cite{sp} it was shown that it is possible to have p- and S-brane intersections without this assumption and 
these two types were called as {\it Class I} and {\it Class II} intersections respectively. In all these works, Chern-Simons terms in the action
do not play any role. In \cite{chern} an exact S-brane configuration in D=11 was found in which Chern-Simons contribution to field equations 
is nonzero. It was found in \cite{deform}  that this solution can also be obtained by applying a Lunin-Maldacena deformation \cite{lm} to a single SM2-brane 
solution using a formula derived in \cite{deformations}.

In this paper our goal is to obtain intersections of S-branes with two other fundamental types of exact solutions, namely
waves (W) and Kaluza-Klein (KK) monopoles. These are purely geometric backgrounds of String/M-theory and their intersections with p-branes were studied 
in \cite{bergmono}. In section 2 we give the S-brane solution in a slightly more general form from the one given in \cite{s2} by 
allowing more integration constants which are required for intersections with waves. We take the transverse space to be flat since that is 
the only allowed option in such intersections. 
In section 3 we study intersections with wave and find that such a 
solution exists only if the wave is located in the transverse space of the
S-brane. This is significantly different from p-branes where the wave is placed inside the p-brane worldvolume \cite{bergmono}.
The effect of adding a wave to an S-brane is just a linear condition on integration constants of time dependent metric functions of the S-brane.
In section 4 we show that no such condition arises in S-brane intersections with KK-monopoles and there are several
ways to place the KK-monopole similar to the p-brane case \cite{bergmono}. We list all possible configurations in D=11 at the end of sections 3 and 4
and determine all intersections between S-branes and D0- or D6-branes that arise in D=10 upon dimensional reduction. They are consistent
with results of \cite{sp} where these solutions were obtained by directly solving field equations. As is well-known, intersecting p-brane solutions play an important 
role in AdS/CFT duality \cite{maldacena, gubser, witten}
and in understanding the black hole (BH) physics from string/M-theory (see \cite{gaunt, kostas} for a review). In section 5 we investigate whether it is 
possible to add an S-brane to these configurations. Such a solution would provide an opportunity to study these objects in a time-dependent background.
However, after systematically analyzing all known D=11 cases we found that this is not allowed and we illustrate this in detail for the M2-M5 case. 
The main difficulty is that adding p-branes puts restrictions on integration constants of the S-brane that can not be satisfied simultaneously.
In the appendix we show that deforming the worldvolume of the S-brane, which increases the 
number of integration constants, does not change our negative result in section 5. However, we find that this may reduce the number of smeared directions in 
intersections with p-branes. We conclude with some comments in section 6.

\section{S-branes}
In this section we will describe the single S-brane solution of supergravity theories given in \cite{s2} by allowing more number of integration constants. 
The general action in $d$-dimensions describing the bosonic sector of various supergravity theories comprising the metric 
$g_{MN}$, a gauge potential $A_{[n-1]}$ with corresponding field strength $F_{[n]}=dA_{[n-1]}$ and a dilaton scalar 
field $\phi$ coupled to the form field with the coupling constant $a$ is given in the Einstein frame as
\be
S=\int d^d x \sqrt{-g} \left( R- \frac{1}{2} \del_{\m}\f \del^{\m} \f - 
\frac{1}{2(n)!}e^{a \f} F^2_{[n]} \right) \, , 
\label{action}
\ee
Here Chern-Simons terms are omitted since they are irrelevant for the solutions that will be considered in this paper.
The field equations are
\bea
&&R_{\m\n}=\frac{1}{2}\del_\m \f \del_\n \f +
\frac{1}{2(n)!}e^{a \f} \left( n F_{\m{\a_2}...{\a_{n}}} F_\n 
^{\, \a_2...\a_{n}} - \frac{(n -1)}{d-2}F^2_{[n]}g_{\m\n} \right) \,\, , \nonumber\\
\nonumber
&&\del_\m \left( \sqrt{-g}e^{a \f} F^{\m \n_2 ...\n_{n}} \right)=0 \,\,\,\, , \\
\label{fieldeqns}
&&\frac{1}{\sqrt{-g}} \del_\m \left( \sqrt{-g}\del^\m \f 
\right)=
\frac{a}{2(n)!}e^{a \f}F^2_{[n]} \,\,\,\, .
\eea
There is also the Bianchi identity $\del _{[\n}F_{\m_1...\m_{n}]}=0$.

The metric for an S(p-1)-brane in $d$-dimensions with a $p$-dimensional Euclidean worldvolume 
and a transverse space comprising of a $k$-dimensional localized flat space and a $q-k+1$ delocalized flat directions is
\be
ds^2=-e^{2A}dt^2+e^{2B}(dx_1^2+...+dx_p^2)+e^{2D}(dy_1^2 + ... + dy_k^2)+  e^{2C}dz^2  + e^{2E} (dw_1^2 + ... + dw_{q-k}^2)
\label{smetric}
\ee
Here all the metric functions are time dependent and $p+q=d-2$. 
Although, in general it is possible to have constant curvature spaces in the transverse part of the S-brane, for our purposes it is enough
to consider a flat space. Moreover,  we singled out the $z$ direction which will be used in the intersection with waves.

The field equations (\ref{fieldeqns}) and Bianchi identities for the field strength of an S(p-1)-brane can be solved as 
\bea
F_{[d-p-1]} &=& Q_s dy_1 \wedge ... dy_k \wedge dz \wedge dw_1 \wedge ... \wedge dw_{q-k}  \hspace{1.2cm} \textrm{(magnetic)} \, , \\
F_{[p+1]} &=& Q_s * (dy_1 \wedge ... dy_k \wedge dz \wedge dw_1 \wedge ... \wedge dw_{q-k}) \hspace{0.5cm} \textrm{(electric)} \, , \nonumber
\label{formfield}
\eea
where the Hodge dual $*$ is with respect to the full metric and $Q_s$ is the charge of the S(p-1)-brane. The Ricci tensor 
for the above metric (\ref{smetric}) largely simplifies with the gauge condition 
\begin{eqnarray}\label{gaugeSbranewave}
-A+pB+C+kD+(q-k)E=0 \, , 
\end{eqnarray}
which corresponds to the time reparametrization invariance. After these, the metric functions and dilaton are found as
\bea
B&=&\frac{2}{\chi}\ln{\left(\frac{m}{Q_s\cosh{\left[m(t-t_0)\right]}}\right)} 
-\frac{a \gamma }{\chi}t - \frac{a \Omega}{\chi} \nonumber , \\
C &=&-\frac{p}{q} B+c_1t +d_1, \nonumber \\
D&=&-\frac{p}{q} B - \left(c_2 + \frac{1}{k-1} [c_1 + (q-k)c_3]\right)t + d_2, \nonumber \\
E &=&-\frac{p}{q} B+c_3t +d_3, \nonumber\\
A &=&-\frac{p}{q} B - \left( kc_2 + \frac{1}{k-1}[c_1 + (q-k)c_3] \right)t + [d_1+kd_2 +(q-k)d_3] , \nonumber \\
\phi &=& - \frac{\epsilon a(d-2)}{q} B + \gamma t + \Omega \, .
\label{standard}
\eea
The constants satisfy
\be
k(k-1)c_2^2 =  \frac{m^2}{2} + [c_1^2+(q-k)c_3^2] + \frac{1}{k-1}[c_1 + (q-k)c_3]^2 + \frac{p}{\chi} \gamma^2 \, ,
\label{m2}
\ee
where
\be
\chi = 2p + \frac{a^2 (d-2)}{q} \, .
\label{dilaton}
\ee
We also have the condition $k \geq 2$. In 11-dimensional supergravity there is no dilaton 
and the dilaton coupling '$a$' is zero. The value of  '$a$' for an S(p-1)-brane 
in type $IIA$ and $IIB$ 
supergravities are given by:
\be
\begin{cases} \epsilon a=\frac{4-p}{2} \, , \hspace{1cm}
\textrm{$RR$-branes} \cr
\epsilon a=\frac{p-4}{2} \, , \hspace{1cm}
\textrm{$NS$-branes}
\end{cases}
\ee
where $\epsilon = 1$ for electric branes ($p=1,3$  in type $IIA$ 
and 
$p=2$ in type 
$IIB$) and $\epsilon=-1$ for magnetic branes ($p=5,7$ in type 
$IIA$ 
and $p=6$ in type $IIB$). In these theories we also have the identity \, $q \chi = 4 (d-2)$.

The above solution is slightly more general than the solution given in \cite{s2} for the flat transverse space where it was assumed that $d_1=d_2=0$ and more
non-trivially $c_1=c_3=0$. Having nonzero $c_1$ and $c_3$ is crucial for intersections with waves as we will see in (\ref{waveconstant}) in
the next section.  Finally, let us note that if
either the S(p-1)-brane charge $Q_s$ or the constant $m$ is zero, then the other one should also be zero and 
we should take $m/Q_s=1$ in the above solution (\ref{standard}). Moreover in the function $B$ in equation (\ref{standard}) the replacement 
$ \ln (\cosh [m(t-t_0)]) \rightarrow \hat{m}(t-t_0)$ should be made. These are also required when $k=1$. Note that, in this limit the metric functions and the 
dilaton become linear in time.

\section{Intersections of S-branes with Plane Waves}
\label{IntSbraneWave}

The plane wave in $d$-dimensions propagating in the $z$-direction is given by the metric
\begin{equation}
ds^2=-e^{-2G}dt^2+e^{2G}{\left[(e^{-2G}-1)dt+dz\right]}^2+ dy_1^2+...+dy_{(d-2)}^2 \, .
\label{wavemetric}
\end{equation}
This is a solution of vacuum Einstein equations of the above action (\ref{action}) provided that $e^{2G}$ is a harmonic function of the variables 
$\{y_1,...,y_{(d-2)}\}.$ Some of these coordinates can also be taken as isometry directions.

We will show that a plane wave can be added only on the transverse space of an S-brane. 
This is essentially due to the fact that by definition the time coordinate is 
transverse to the S-brane which forces the $z$ coordinate of the wave (\ref{wavemetric}) 
also to be in the transverse space of the S-brane. The harmonic function of the wave depends 
on some of the spatial transverse coordinates of the S-brane. In general in S-brane solutions it is allowed to 
have a spherical or a hyperbolic space on its transverse space (\ref{smetric}). 
However, in the presence of a wave this is not possible and the transverse space must be flat for the 
separability of the time dependent functions and the spatial function of the wave. This is actually expected, since
when the S-brane is removed from the configuration by setting its charge to zero, transverse part of the wave needs to be flat.
We write the metric that describes such an intersection in $d$-dimensions as  
\bea
ds^2=&-&e^{2A-2G}dt^2+e^{2B}({dx}{^{2}_{1}}+...+{dx}{^{2}_{p}})+e^{2C+2G}\left[(e^{-2G}-1)e^{A-C}dt+dz\right]^2 \nonumber \\
&+& e^{2D}({dy}{^{2}_{1}}+...+{dy}{^{2}_{k}})+e^{2E}(dw_1^2+...+dw_{(q-k)}^2) \, . 
\label{swavemetric}
\eea
The metric above is parametrized by five time dependent functions $A(t),B(t),C(t),D(t)$ and $E(t)$ and the wave function $G=G(y_a)$. 
Note that when $G=0$ the metric reduces to the S-brane metric (\ref{smetric}). 
With the coordinates $x^A=(t,x^{\mu},z,y^a,w^i)$ the vielbein reads:
$$
E^{\bar{t}}=e^{A-G}dt, E^{\bar{\mu}}=e^{B}dx^\mu, E^{\bar{a}}=e^{D}dy^a, E^{\bar{i}}=e^{E}dw^i, E^{\bar{z}}=(e^{A-G}-e^{A+G})dt+e^{C+G}dz .
$$
In Ricci tensor components of this metric several terms of the form $[\partial_a\partial_aG+2(\partial_aG)(\partial_aG)]$ 
appear which all vanish as one imposes $e^{2G(y_a)}$ to be a harmonic function, that is,
\begin{eqnarray}
\partial_a\partial_ae^{2G}=0.
\end{eqnarray}
The remaining field equations are exactly those of an S-brane except that one has to account 
for the new Ricci tensor components brought about by the wave function
\begin{eqnarray}
\label{non1}
R_{ta}&=&e^{2G}(\partial_aG)\left[-p\dot{B}-(q-k)\dot{E}-2\dot{C}-(k-2)\dot{D}\right],\\
\label{non2}
R_{za}&=&e^{2G+C-A}(\partial_aG)\left[p\dot{B}+(q-k)\dot{E}+2\dot{C}+(k-2)\dot{D}\right] \, ,
\end{eqnarray}
where dot represents the derivative with respect to time. Since in the metric (\ref{swavemetric}) neither $t$ nor $z$ mix with the $y_a$'s and since the dilaton 
and the gauge field depend only on time, these Ricci components must vanish on their own. Using the gauge condition 
equation (\ref{gaugeSbranewave}) in (\ref{non1}) and (\ref{non2}), it is seen that both these components vanish if
\begin{eqnarray}\label{wavecondgen}
\dot{A}+\dot{C}=2\dot{D}
\end{eqnarray}
which we are going to call the {\it wave condition}. Note that $A, C, D$ are time dependent metric functions corresponding respectively to time, wave propagation
direction and the space where the wave function depends. Since the remaining field equations are exactly those of an S-brane, we 
have the usual S-brane solution (\ref{standard}) together with the wave condition (\ref{wavecondgen}) which gives a relationship
between integration constants in functions $A,C$ and $D$ corresponding to linear terms in time as
\be
kc_1 + (q-k)c_3 - (k-1)(k-2)c_2 =0 \, .
\label{waveconstant}
\ee
Now let us focus on D=11 where for SM5 $p=6,~q=3$ and for 
SM2 we have $p=3,~q=6$. The metric functions in (\ref{swavemetric}) are given in (\ref{standard}) with $a=0$.
In the orthonormal frame the 4-form field strengths for SM5 and SM2 with charge parameter $Q_s$ are as follows:
\begin{eqnarray}
F_{SM5}&=& Q_se^{6B-A+G}E^{\bar{z}}\wedge E^{\bar{y}_1}\wedge...\wedge E^{\bar{y}_k}\wedge E^{\bar{w}_1}\wedge...\wedge E^{\bar{w}_{(3-k)}} \, , \\
F_{SM2}&=& Q_se^{3B-A+G}E^{\bar{t}}\wedge E^{\bar{x}_1}\wedge E^{\bar{x}_2}\wedge E^{\bar{x}_3}.
\end{eqnarray}
We summarize these intersections below using the convention employed in \cite{bergmono} for p-branes
where a worldvolume direction is denoted by a '$\times$' symbol and a transverse direction by a '$-$' symbol. 
The portion of the metric that involves the term $e^{2G}{\left[(e^{-2G}-1)dt+dz\right]}^2$ is denoted by the $z$ coordinate in the table:

\begin{tabular}{|c||c|c|c|c|c|c|c|c|c|c|c|}
\hline
$SM5$ &$t$ &$\times$ &$\times$ &$\times$ &$\times$ &$\times$ &$\times$ &$-$ &$-$ &$-$ &$-$ \\ \hline
$W$ &$t$ &$-$ &$-$ &$-$ &$-$ &$-$ &$-$ &$z$ &$-$ &$-$ &$-$ \\ \hline
\hline
$SM2$ &$t$ &$\times$ &$\times$ &$\times$ &$-$ &$-$ &$-$ &$-$ &$-$ &$-$ &$-$ \\ \hline
$W$ &$t$ &$-$ &$-$ &$-$ &$z$ &$-$ &$-$ &$-$ &$-$ &$-$ &$-$ \\ \hline 
\end{tabular}

\

From these configurations we can obtain various solutions in D=10 by dimensional reduction using the reduction rules for waves \cite{bergmono} and 
S-branes \cite{roy}:

\noindent Wave $\xrightarrow{\rm{\,\, transverse}}$ \,\,\, Wave \, , \, Wave $\xrightarrow{\hspace{0.5cm}\rm{z} \hspace{0.5cm}}$ \,\,\, D0

\noindent SM2 $\xrightarrow{\rm{\,\, transverse}}$ \,\,\, SD2 \, , \, SM2 $\xrightarrow{\rm{\, \, worldvolume}}$ \,\,\, SNS1

\noindent SM5 $\xrightarrow{\rm{\,\, transverse}}$ SNS5 \, , \, SM5 $\xrightarrow{\rm{worldvolume}}$ SD4

\noindent Intersections between S- and p-branes can be classified as {\it Class I} \cite{mas} and {\it Class II} \cite{sp} depending on whether the worldvolume of the S-brane includes the radial part
of the p-brane or not, respectively. Reducing the above D=11 solutions along the $z$-coordinate 
we obtain $(0|D0,SNS5|6)$ and $(0|D0,SD2|3)$ solutions which belong to {\it Class II}. 
\footnote{In the notation $(I|Dp, Sq|b)$ the integer $I$ refers to the number of intersecting dimensions and $b$ is the minimum number of 
smeared directions for the Dp-brane if Sq brane is removed from the solution by setting its charge to zero.} We see that 
there is no solution with a D0-brane in {\it Class I} as was found in \cite{sp}.

\section{Intersections of S-branes with KK Monopoles}
\label{IntSbraneKK}

Kaluza-Klein monopole is another purely geometric solution of the above action (\ref{action}) and its metric 
in $d$-dimensions reads, for $i=1,2,3$
\begin{eqnarray}
ds^2=-dt^2+dx_1^2+...+dx_{(d-5)}^2+e^{-2M}{(dz+J_idy_i)}^2+e^{2M}(dy_1^2+dy_2^2+dy_3^2) \, .
\label{mon1}
\end{eqnarray}
where $e^{2M}$ and $J_i$ depend on $\{y_i\}$, $e^{2M}$ is a harmonic function and it is related to $J_i$ as
\begin{eqnarray}
{\hat F}_{ij}\equiv \partial_i J_j-\partial_j J_i=\epsilon_{ijk}\partial_k e^{2M} \, .
\label{mon2}
\end{eqnarray}
A special case of the given configuration is when one of the $y_i$'s corresponds to an isometry direction. 
Suppose in (\ref{mon1}) and (\ref{mon2}) $y_1$ is taken as an isometry direction. Then equation (\ref{mon2}) becomes
\begin{eqnarray}
\partial_2J_3-\partial_3J_2&=&0 \label{mon3} \, , \nonumber\\
\partial_3J_1&=&\partial_2e^{2M} \, ,\nonumber \\
-\partial_2J_1&=&\partial_3e^{2M} \, .
\end{eqnarray}
Furthermore, in the Riemann tensor $J_2$ and $J_3$ appear only in the combination $(\partial_2J_3-\partial_3J_2)$ 
which by equation (\ref{mon3}) is equal to zero. This means that they can be gauged away. In this case one should no 
longer assume spherical symmetry in the $\{y_i\}$ space and observe that both $e^{2M}$ and $J_1$ are harmonic in $y_2$, $y_3$.

Several configurations are possible for intersections of an S-brane and a Kaluza-Klein monopole. 
Since the KK-metric is Ricci flat, placing it 
in the worldvolume or the flat transverse space of an S-brane as a whole is guaranteed to be a solution. It turns out
that one may also put only part of it to S-brane's worldvolume without altering other fields than the metric.

We will summarize all possible configurations between a single S-brane and a KK-monopole in D=11 in the table below.
We denote the portion of the metric that involves the term $e^{-2M}(dz+J_idy_i)^2$ by the $z$ coordinate of 
the Kaluza-Klein monopole and each of the $y_i$ directions, which appear with the line element $e^{2M}dy_i^2$ in the metric, 
by the corresponding $J_i$. We call these $\{z,J_i\}$ directions transverse to KK and the rest of the coordinates as 
"worldvolume" or "brane" directions of the KK. In these solutions when the full Kaluza-Klein metric with no additional isometry implemented, 
the $\{z,J_i\}$ part of the metric must have the same overall time dependent factor. However, when an isometry direction is assumed, say $y_1$, 
one can write the metric with different time factors multiplying the $\{z,J_1\}$ and $\{J_2,J_3\}$ parts.

\

\begin{tabular}{|c||c|c|c|c|c|c|c|c|c|c|c|}
\hline
$SM5$ &$t$ &$\times$ &$\times$ &$\times$ &$\times$ &$\times$ &$\times$ &$-$ &$-$ &$-$ &$-$ \\ \hline
$KK$ &$t$ &$J_1$ &$J_2$ &$J_3$ &$z$ &$\times$ &$\times$ &$\times$ &$\times$ &$\times$ &$\times$ \\ \hline  
\hline
$SM5$ &$t$ &$-$ &$\times$ &$\times$ &$-$ &$\times$ &$\times$ &$\times$ &$\times$ &$-$ &$-$ \\ \hline
$KK$ &$t$ &$J_1$ &$J_2$ &$J_3$ &$z$ &$\times$ &$\times$ &$\times$ &$\times$ &$\times$ &$\times$ \\ \hline
\hline
$SM5$ &$t$ &$\times$ &$-$ &$-$ &$\times$ &$\times$ &$\times$ &$\times$ &$\times$ &$-$ &$-$ \\ \hline
$KK$ &$t$ &$J_1$ &$J_2$ &$J_3$ &$z$ &$\times$ &$\times$ &$\times$ &$\times$ &$\times$ &$\times$ \\ \hline
\hline
$SM5$ &$t$ &$-$ &$-$ &$-$ &$-$ &$\times$ &$\times$ &$\times$ &$\times$ &$\times$ &$\times$ \\ \hline
$KK$ &$t$ &$J_1$ &$J_2$ &$J_3$ &$z$ &$\times$ &$\times$ &$\times$ &$\times$ &$\times$ &$\times$ \\ \hline
\hline
$SM2$ &$t$ &$-$ &$\times$ &$\times$ &$-$ &$\times$ &$-$ &$-$ &$-$ &$-$ &$-$ \\ \hline
$KK$ &$t$ &$J_1$ &$J_2$ &$J_3$ &$z$ &$\times$ &$\times$ &$\times$ &$\times$ &$\times$ &$\times$ \\ \hline
\hline
$SM2$ &$t$ &$-$ &$-$ &$-$ &$-$ &$\times$ &$\times$ &$\times$ &$-$ &$-$ &$-$ \\ \hline
$KK$ &$t$ &$J_1$ &$J_2$ &$J_3$ &$z$ &$\times$ &$\times$ &$\times$ &$\times$ &$\times$ &$\times$ \\ \hline
\hline
$SM2$ &$t$ &$\times$ &$-$ &$-$ &$\times$ &$\times$ &$-$ &$-$ &$-$ &$-$ &$-$ \\ \hline
$KK$ &$t$ &$J_1$ &$J_2$ &$J_3$ &$z$ &$\times$ &$\times$ &$\times$ &$\times$ &$\times$ &$\times$ \\ \hline
\end{tabular}

\

We write the metric and the form field explicitly for two of the given SM5 configurations for clarity. 
For the first SM5-brane configuration the metric reads
\bea
ds^2=&-&e^{2A}dt^2+e^{2B}\left[\delta _{\mu\nu}dx^{\mu}dx^{\nu}+
e^{-2M}{(dz+J_idy_i)}^2+e^{2M}(dy_1^2 + dy_2^2 +dy_3^2)\right] \nonumber \\&+&e^{2C}\delta_{mn} dw^mdw^n +e^{2D}\delta _{ab}dv^adv^b \, ,
\eea
where $\mu=1,2$; $i=1,2,3$; $m=1,...,k$; $a=1,...,(4-k)$; $k>1$. The 4-form field is exactly the same as it is with a single SM5-brane (\ref{formfield}) and
the metric functions can be read of from  (\ref{standard}).

The metric of the third SM5-brane configuration reads
\bea
ds^2=&-&e^{2A}dt^2+e^{2B}\left[\delta _{\mu\nu}dx^{\mu}dx^{\nu}+
e^{-2M}{(dz+J_1dy_1)}^2+e^{2M}(dy_1)^2\right]\nonumber \\&+&e^{2C+2M}\left[dy_2^2 + dy_3^2\right]+e^{2D}(dw_1^2 +dw_2^2) \, ,
\eea
where $\mu=1,...,4$. For this configuration the 4-form field is
\begin{eqnarray}
F_{[4]}=Q_s e^{2M} dy_1 \wedge dy_2 \wedge dw_1 \wedge dw_2 \, .
\end{eqnarray}

Note that in the fourth configuration the Kaluza-Klein monopole is entirely in the transverse space of the SM5-brane.
The second SM5 configuration is similar to the third.

We can again do dimensional reduction to D=10 using the following rule for the KK monopole \cite{bergmono}:

\noindent KK $\xrightarrow{\rm{ \,\, transverse \, \, z}}$ D6 \, , \, KK $\xrightarrow{\rm{\,\, worldvolume \, \, x^{\mu}}}$ KK  

\noindent Here the D6-brane is located at worldvolume
directions of the KK-monopole which are indicated with the symbol '$\times$' in KK rows in the table.
Upon reduction along the $z$-coordinate to D=10 from the first two SM5-KK configurations we obtain $(2|D6,SD4|0)$ and $(4|D6,SNS5|1)$ solutions respectively
which are in {\it Class I}, i.e. the radial coordinate of the D6-brane is included in the worldvolume of the S-brane. Whereas, the last two SM5-KK
configurations give rise to $(4|D6,SD4|1)$ and $(6|D6,SNS5|0)$ solutions respectively, that belong to {\it Class II}. Similarly, from the first SM2-KK intersection
we get  $(1|D6,SD2|1)$ which is in {\it Class I}. From the last two SM2-KK intersections we get $(3|D6,SD2|0)$ and $(1|D6,SNS1|1)$ solutions of {\it Class II} 
respectively. \footnote{$(1|D6,SNS1|1)$ solution was not listed explicitly in \cite{sp} since here the transverse space of the D6-brane is only 2-dimensional which results in 
a logarithmic harmonic function.}
We also see that D6-SNS1 intersection is not possible in {\it Class I} as was observed in \cite{sp}.

\section{S-Branes and AdS Backgrounds}

Intersecting brane solutions with waves and KK-monopoles in $D=11$ and $D=10$ play an important role in the study of AdS/CFT duality \cite{maldacena, gubser, witten} and 
give a higher dimensional viewpoint to some 
black holes in four and five dimensions \cite{gaunt, kostas}. Since we now have intersections of S-branes with waves and monopoles and
since intersections of S-branes with p-branes are already known \cite{mas, sp} we are ready to investigate if it is possible
to add an S-brane to these configurations. If it were, it would provide an opportunity to study AdS/CFT duality and lower dimensional black holes
in a time dependent background. We focus on D=11 configurations since D=10 cases can be obtained from them by dimensional reduction and 
T and S-dualities. Metric of such an intersection is written in factorized form as:
\be
ds^2=-e^{2A(t)} e^{2\a(r)} dt^2\,+\,\sum_q\,e^{2C_q(t)} e^{2\b_q(r)}\,ds_q^2\,+\,e^{2D(t)} e^{2\theta(r)}\,(dr^2 + r^2d\Omega_m^2) \, ,
\ee
where $ds_q$ is the metric on the $d_q$ dimensional flat space. From this metric we get the following non-diagonal Ricci tensor component:
\be 
R_{tr} = \left[m\dot{D}\a' + \sum_q d_q\left(\dot{C}_q\a' + [\dot{D} -\dot{C}_q]\b_q'\right)\right] \, ,
\label{mix}
\ee
where prime indicates differentiation with respect to $r$.
Meanwhile, assuming that the field strength is summation of the corresponding field strengths of S- and p-branes with appropriate
$r$ and $t$ dependent multiplicative metric functions respectively,
radial and time-dependent parts decouple and we end up with the usual S and p-brane equations \cite{sp} in addition to (\ref{mix}). 
By solving (\ref{mix}) one fixes the intersection dimension and minimum amount of smearing that is necessary for the p-brane.
From the field equations (\ref{fieldeqns}) we see that in D=11, equation (\ref{mix}) should vanish on its own and we find \cite{sp}:
\bea
&& {\bf Class \, I \,:} (0|M5, SM2|2), (2|M5, SM5|1), (0|M2, SM5|2)   \textrm{(radial part inside S-brane)} \nonumber \\
&& {\bf Class \, II :} (2|M5, SM2|1), (4|M5, SM5|2), (2|M2, SM5|5), (1|M2, SM2|2)   
\label{list}
\eea 
In the last three solutions of {\it Class II}  in (\ref{list}) the S-brane has zero charge. 
That is why they were not listed in the table in \cite{sp} although their existence follows from the intersection rule found in \cite{sp}. 
However, in the last one if we allow more than 
2 smearings for the M2-brane, such as $(1|M2, SM2|3)$, it is possible to have a charged SM2. In the other two chargeless cases, 
there is no room for extra smearing if we insist on having at least 3 dimensional localized transverse space for the M-brane. 
Finally, one can have 
these three configurations with a charged S-brane if one allows non-homogeneous worldvolume for the S-brane as we shall show in the appendix.

From the list (\ref{list}) we see that M-brane and SM-brane intersections are allowed only with
smearing for the M-brane and hence it is not possible to have $AdS_7$ and $AdS_4$ near horizon geometries
in the presence of an S-brane. However, there are also M-brane intersections with AdS near-horizon geometries \cite{ads} which describe 
four or five dimensional black holes  \cite{black1, black2}: 

\noindent D=5 BH: a) M2-M2-M2 $\xrightarrow{{\rm near \,\, horizon}}$ $AdS_2 \times S^3 \times E^6$ \\
\phantom{D=5 BH:} b) M2-M5  $\xrightarrow{{\rm near \,\, horizon}}$ $AdS_3 \times S^3 \times E^5$

\noindent D=4 BH: a) M2-M2-M5-M5  $\xrightarrow{{\rm near \,\, horizon}}$ $AdS_2 \times S^2 \times E^7$ \\
\phantom{D=4 BH:} b) M5-M5-M5 $\xrightarrow{{\rm near \,\, horizon}}$ $AdS_3 \times S^2 \times E^6$

\noindent where $E^l$ is the $l$-dimensional Euclidean space. M2-M5 and M5-M5-M5 solutions intersect on a common string to which a
wave should be added to obtain a black hole with non-vanishing horizon area. From (\ref{list}) at first sight it seems possible to add 
an S-brane to these configurations without any extra smearing.\footnote{In the above intersections with two M5-branes we have M5-M5(3),
i.e. they intersect on a 3-dimensional space. There is another supersymmetric case, namely M5-M5(1) where the harmonic functions depend on relative
transverse directions and we have: M5-M5(1)-M2 $\xrightarrow{{\rm near \,\, horizon}}$ $AdS_3 \times S^3 \times S^3 \times E^2$.

There is also an M2-M5 intersection where M2-brane is contained in the worldvolume of the M5-brane together with a static SM2-brane \cite{dyon1, dyon2}
and we have:
M2 $\subset$ M5 $\xrightarrow{{\rm near \,\, horizon}}$ $AdS_7 \times S^4$. 

In both of these cases, when we try to add an S-brane
extra smearings are needed (\ref{list}) which ruin the AdS geometry.}
However, we found that 
this is not allowed in any of the above cases. This is because each M-brane imposes a condition
on integration constants in solving (\ref{mix}) and these do not work with the S-brane condition (\ref{m2}).
Now we will illustrate this for the M2-M5 intersection in detail.

\

{\bf M2-M5 Intersection}

From the list (\ref{list}) we see that we can add an SM5 from {\it Class I} and SM2 or SM5 from {\it Class II} to this intersection. 
However, one can easily check that SM5 of {\it Class II} requires an additional smearing and hence spoils the AdS geometry. Moreover,
SM5 of {\it Class I} does not leave any room for the wave function to depend. So, it is enough to focus on SM2 from {\it Class II}.
The unique metric of the SM2-M2-M5-W configuration following intersection rules above (\ref{list}) with no extra smearing is:

\bea
ds_{11}^2 &=&H_{M5}^{-1/3}\Big(H_{M2}^{-2/3}\big\{-e^{2A-2G} dt^2+e^{2C+2G} {\left[(e^{-2G}-1)e^{A-C}dt+dz\right]}^2\big\} 
\nonumber \\ 
&+& H_{M2}^{1/3}\big\{e^{2B}(dx_1^2+dx_2^2)+e^{2E}(dw_1^2+dw_2^2)\big\}\Big) +  H_{M5}^{2/3}H_{M2}^{-2/3} e^{2B} dx_3^2 \nonumber \\
&+& H_{M5}^{2/3}H_{M2}^{1/3}e^{2D}(dr^2+r^2d\Omega _3^2) \, ,
\label{m2m5metric}
\eea
where the $\{x_1,x_2, x_3\}$-space describes the worldvolume of 
the SM2-brane. M2-brane is located at $\{t,z, x_3\}$ and M5-brane is located at $\{t,z,x_1,x_2,w_1, w_2\}$.
Note that the position of the wave is consistent with our findings in section 3.
The field strengths are added up with appropriate $r$ and $t$ dependent modifications \cite{sp}. 
Harmonic functions are taken to be isotropic in the four dimensional overall transverse space:
\be
H_{M2}=1+\frac{Q_{M2}}{r^2}~,~~~H_{M5}=1+\frac{Q_{M5}}{r^2}~,~~~e^{2G}=1+\frac{Q_W}{r^2} \, .
\ee

Starting with the SM2-brane solution given in (\ref{standard}) with $p=3, q=6$ and $k=4$, equation (\ref{mix}) brings (assuming independent charges),
\begin{eqnarray}
\textrm {M2:} && \dot{D}+\dot{B}+\dot{E}=0 \, , \nonumber\\
\textrm{M5:} && 2\dot{D} + \dot{B} =0 \, . 
\label{m2m5}
\end{eqnarray}
Using these in (\ref{standard}) we see that
$c_3=0$ and $3c_2= -c_1$. Now, (\ref{m2}) implies that $m=0$  which enforces the charge of SM2, i.e.  $Q_s$, to be zero.
But the spacetime where the common string lives still has time dependence through the constant $c_1$, i.e. $A=C=c_1t$. However, this 
time dependence is trivial since by defining null-coordinates $u=t+z$ and $v=t-z$ and making coordinate changes $d\tilde{u}= e^{c_1u}du$
and $d\tilde{v}= e^{c_1v}dv$, it can be removed from the metric and from the 4-form field strength of M2-brane which initially has a multiplicative factor
of $e^{2c_1t}$. Now to add a wave we should impose the wave condition (\ref{wavecondgen}),
\be
\dot{A}+\dot{C}-2\dot{D}=0  ~~~~ \Rightarrow ~~~~ 2 c_1 = 0 ,
\label{WaveconS1d5W}
\ee
after which what remains is just the M2-M5-W solution.\footnote{However, instead of adding a wave we can use coordinates $\{z, x_1\}$ and a coordinate from the overall 
transverse direction to add 
an SM2-brane with a non-homogeneous worldvolume along these directions (see the appendix) using the solution generating technique described in \cite{deform}. Since, there is an additional 
smearing, AdS geometry disappears.} Since the difficulty arises due to conditions on integration constants, increasing their number looks promising. 
For that reason, in the appendix we deform the worldvolume of the SM2-brane that leads to two additional integration constants which, still does not change 
the no-go result. Finally, adding another SM2-brane along $\{w_1, w_2, x_3 \}$ directions does not help either.

\section{Conclusions}

S-branes have many fundamental applications in String/M-theory and hence it is desirable to increase available S-brane solutions. 
With this main motivation, in this paper we constructed intersections of S-branes with waves and monopoles which are among the most important
exact solutions of String/M-theory. This completes a big gap in our knowledge about S-brane solutions. Moreover, as we saw at the end of sections 3 and 4, 
these solutions give the 11-dimensional origin of the intersections between
S-branes with D0 and D6-branes in D=10. There is no intersection between a wave and monopole along a common $z$-direction 
\cite{bergmono} and hence we covered all possibilities.

Having all intersections between S-branes and basic static solutions of String/M-theory at hand,
we next looked at the problem of adding an S-brane to M-brane 
intersections with AdS near horizons in section 5.  We found that this is not possible. It is certainly desirable to look for
ways of doing this. It seems that some radical changes in our approach are needed.
For example, instead of orthogonal intersections of S-branes with BPS p-branes one may allow intersections 
with angles and intersections with black p-branes \cite{pblack1, pblack2}. One may also consider 
adding S-branes to partially localized solutions, such as M5-M5(3) where harmonic function of one of the p-brane 
depends other's worldvolume directions in addition to overall transverse coordinates\cite{localized} .

In the M- and SM-brane intersections, the S-brane worldvolume 
directions are multiplied with different radial functions. 
Upon compactification, SM2-branes give rise to accelerating 4-dimensional 
cosmologies \cite{townsend}-\cite{present}. Having different radial dependence may be useful in studying non-homogeneous cosmologies.

Finally, there is another class of time 
dependent p-brane solutions called {\it dynamical branes} where 
the harmonic function of the p-brane is modified by adding a linear time dependent term \cite{dynamic1, dynamic2}.
It would be interesting to study intersections between these and S-branes. We hope to explore these problems in near future.

\

\noindent {\Large \bf Acknowledgements}

\noindent NSD is partially supported by Tubitak grant 113F034.

\appendix

\section{Non-homogeneous S-branes}

In this appendix we allow the worldvolume of the S-brane to be non-homogeneous \cite{ohta1} to increase the number of integration constants in the solution 
(\ref{standard}). It is possible to deform the worldvolume of the S-brane given in (\ref{smetric}) without altering the other metric functions  and 
physical fields given in (\ref{standard}) as follows:
$$
e^{2B}(dx_1^2+...+dx_p^2) \rightarrow e^{2B}e^{-2m/p}(e^{2b_1t}dx_1^2+...+ e^{2b_pt}dx_p^2)
$$
where $b_i$'s are constants that satisfy $\displaystyle \sum_{i=1}^{p} b_i=m $. The condition (\ref{m2}) should be modified as
\be
k(k-1)c_2^2 =  \frac{(p-2)}{2p}m^2 + \sum_{i=1}^{p} (b_i)^2 + [c_1^2+(q-k)c_3^2] + \frac{1}{k-1}[c_1 + (q-k)c_3]^2 + \frac{p}{\chi} \gamma^2 \, .
\label{m2new}
\ee
Note that when all $b_i$'s are equal
we have the homogeneous case. If we do this deformation in the M2-M5 intersection (\ref{m2m5metric}) that we considered in section 5,  equation (\ref{m2m5})
should be replaced with
\begin{eqnarray}
\textrm {M2:} && 2\dot{D}+\dot{B_1}+ \dot{B_2}+ 2\dot{E}=0 \, , \nonumber\\
\textrm{M5:} && 2\dot{D} + \dot{B_3} =0 \, , 
\label{m2m5new}
\end{eqnarray}
where $B_i=B+b_i-\frac{m}{3}$ for $i=1,2,3$. From these, it follows that $2c_1=-6c_2-c_3$ which, when used in (\ref{m2new}) implies $c_3=m=b_1=b_2=b_3=0$. 
Hence, no non-trivial time dependence remains.

Now, if we remove M5-brane from the system and take $E=D$ in (\ref{m2m5metric}) to prevent more than 2 smearings for the M2-brane, only the first condition in
(\ref{m2m5new}) remains. In the homogeneous case this implies $m=0$, i.e., a chargeless S-brane. However, we can have a charged S-brane when its worldvolume
is non-homogeneous, by allowing a non-zero $b_1+b_2$.

\end{document}